\documentclass[reprint,amsmath,amssymb,aps,prd,nofootinbib,superscriptaddress]{revtex4-1}
\usepackage{graphicx}
\usepackage{color}
\usepackage{epstopdf}
\usepackage[usenames,dvipsnames]{xcolor}
\usepackage{dcolumn}
\usepackage{bm}
\usepackage{paralist}
\usepackage{amsmath}
\usepackage{pstool}
\usepackage[percent]{overpic}
\usepackage{rotating}

\definecolor{Xred}{HTML}{B31B1B}

\usepackage{scalerel}
\usepackage{tikz}
\usetikzlibrary{svg.path}

\definecolor{orcidlogocol}{HTML}{A6CE39}
\tikzset{
  orcidlogo/.pic={
    \fill[orcidlogocol] svg{M256,128c0,70.7-57.3,128-128,128C57.3,256,0,198.7,0,128C0,57.3,57.3,0,128,0C198.7,0,256,57.3,256,128z};
    \fill[white] svg{M86.3,186.2H70.9V79.1h15.4v48.4V186.2z}
                 svg{M108.9,79.1h41.6c39.6,0,57,28.3,57,53.6c0,27.5-21.5,53.6-56.8,53.6h-41.8V79.1z M124.3,172.4h24.5c34.9,0,42.9-26.5,42.9-39.7c0-21.5-13.7-39.7-43.7-39.7h-23.7V172.4z}
                 svg{M88.7,56.8c0,5.5-4.5,10.1-10.1,10.1c-5.6,0-10.1-4.6-10.1-10.1c0-5.6,4.5-10.1,10.1-10.1C84.2,46.7,88.7,51.3,88.7,56.8z};
  }
}

\newcommand\orcidicon[1]{\href{https://orcid.org/#1}{\mbox{\scalerel*{
\begin{tikzpicture}[yscale=-1,transform shape]
\pic{orcidlogo};
\end{tikzpicture}
}{|}}}}

\usepackage[colorlinks=true,linkcolor=Magenta,citecolor=Magenta,breaklinks=true]{hyperref}
\def\ket#1{\mathinner{|{#1}\rangle}}

\begin{document}

\title{Anomalous spectral lines and relic quantum nonequilibrium}

\author{Nicolas G. Underwood\,\orcidicon{0000-0003-4803-2629}
\hspace{-10pt}
\href{https://arxiv.org/a/0000-0003-4803-2629}{\resizebox{19pt}{!}{
\colorbox{Xred}{\color{white}\textsf{arXiv}}}}\,}
\email{nicolas.underwood@wwu.edu}
\affiliation{Department of Physics and Astronomy, Western Washington University, Bellingham, Washington 98225, USA}
\author{Antony Valentini\,\orcidicon{0000-0003-4131-1385}
\hspace{-10pt}
\href{https://arxiv.org/a/0000-0003-4131-1385}{\resizebox{19pt}{!}{
\colorbox{Xred}{\color{white}\textsf{arXiv}}}}\,}
\email{antonyv@clemson.edu}
\affiliation{Department of Physics and Astronomy, Clemson University, Clemson, South Carolina 29634, USA}
\affiliation{Augustus College, 14 Augustus Road, London SW19 6LN, United Kingdom}
\date{\today}

\begin{abstract}
We describe features that could be observed in the line spectra of relic cosmological particles should quantum nonequilibrium be preserved in their statistics. According to our arguments, these features would represent a significant departure from those of a conventional origin. Among other features, we find a possible spectral broadening that is proportional to the energy resolution of the recording telescope (and so could be much larger than any conventional broadening). Notably, for a range of possible initial conditions we find the possibility of spectral line ``narrowing,'' whereby a telescope could observe a line that is narrower than it is conventionally able to resolve. We discuss implications for the indirect search for dark matter, with particular reference to some recent controversial spectral lines. 
\end{abstract}

\maketitle

%
\section{Introduction}\label{sec:1}

De Broglie-Bohm pilot-wave theory (deBB) \cite{deB28,BV09,B52a,B52b,Holl93} introduces extra so-called ``hidden'' variables to textbook quantum physics in order to fully characterize the state of a quantum system. 
If these variables happen to be distributed according to ``quantum equilibrium'' then the Born rule of quantum probabilities is recovered, and deBB replicates the predictions of textbook quantum physics.
If the variables are not distributed in quantum equilibrium, a dynamical, statistical relaxation has been shown to take place \cite{AV91a,AV92,AV01,VW05,SC12,EC06,TRV12,ACV14,Uthesis} so that quantum equilibrium arises in much the same way as thermal equilibrium does in classical physics. 
The demonstration of this ``quantum relaxation'' has eliminated the need to postulate equilibrium outright, and has prompted interest in possible observable consequences of ``quantum nonequilibrium'' in contemporary experimentation \cite{AV07,AV08,AV09,AV10,CV13,UV15,AV01,CV15,VPV19}.
Quantum nonequilibrium is by definition empirically distinct from textbook quantum theory, and so provides a means by which deBB may be distinguished from both textbook quantum theory and other proposals to explain standard quantum phenomena.\footnote{Among the major formulations of quantum physics, the only two that are empirically distinct from textbook quantum theory are deBB (assuming quantum nonequilibrium) and collapse models. Collapse models do allow for the modification of spectral line shapes \cite{BBU14}, as well as various other effects. See Refs.\ \cite{Bassi03,Bassi12} for in-depth reviews.}

It has been conjectured that the Universe could have begun in a state of quantum nonequilibrium \cite{AV91a,AV92,AV91b,AV96,AV01,AV02A}, or that exotic gravitational effects may even generate nonequilibrium \cite{AV10,AV07,AV04b,AV14}.
Although relaxation may be impeded or prevented for simple quantum states \cite{NU18,ACV14,KV16}, it appears to occur remarkably quickly for sufficiently interacting systems \cite{VW05,TRV12,CS10}.
Hence it is expected that in all but the most exceptional circumstances, any quantum nonequilibrium that was present in the early Universe will have subsequently decayed away \cite{AV91a,AV91b,AV92,AV96}.
On this view the Universe may be understood to have already undergone a subquantum analogue of the classical heat death \cite{AV92,AV01,AV91b}. Nevertheless, data from the cosmic microwave background (CMB) do provide a possible hint for the past existence of quantum nonequilibrium, in the form of a primordial power deficit at large scales (as initially reported \cite{Planck13} and subsequently confirmed \cite{Planck15} by the Planck team). It has been argued that such a deficit is a natural prediction of deBB theory \cite{AV10}, though of course the observed deficit may be caused by something else (or be a mere statistical fluctuation as some argue). Recently, predictions regarding the shape of the power deficit have been made \cite{CV15} and have been compared with CMB data \cite{VPV19}. Primordial quantum nonequilibrium also offers a single mechanism that could account for both the CMB power deficit and the CMB statistical anisotropies \cite{AV15}. At present these are of course only hints.

Potentially it is also possible that quantum nonequilibrium may have been preserved for some species of minimally interacting relic particle \cite{AV01,AV07,AV08,UV15}.
If, for instance, a relic particle species decoupled sufficiently early in the primordial Universe, and if it were sufficiently minimally interacting thereafter, there is a possibility that it may still retain nonequilibrium statistics to this day. 
The purpose of a previous paper \cite{UV15} was to explore various means by which quantum nonequilibrium could be conserved in particles, and examples were given both in terms of inflaton decay as well as relic vacuum modes. 
An assessment of the likelihood of such scenarios requires, however, knowledge of assorted unknown contributing factors--the properties of the specific relic particle species, the correct primordial cosmology, and the extent of the speculated initial nonequilibrium. 
Even so, an actual detection of relic nonequilibrium could occur if it were known to leave telltale experimental signatures.
The purpose of this paper is to follow up on Ref.\ \cite{UV15} by describing such telltale signatures in a fitting context.

As discussed in \cite{UV15}, it is natural to consider signatures of relic nonequilibrium in the context of the indirect search for dark matter. For the reasons described in Sec.~\ref{sec:2} we focus on the search for a ``smoking-gun'' spectral line. 
We present a field-theoretical model of spectral measurement intended to act as an analogue of a telescopic photon detector [for example a calorimeter or charge-coupled device (CCD)]. While the model is admittedly simple, and is certainly not a realistic representation of an actual instrument, it does capture some key characteristics and demonstrates the essential difference between conventional spectral effects and those caused by quantum nonequilibrium. 

Due to the exotic nature of quantum nonequilibrium, the spectral effects we describe are something of a departure from those of a classical origin. For instance, according to our arguments the amount of spectral line broadening depends on the energy resolution of the telescope's photon detector.
Lines may acquire double or triple bumps, or as we discuss, more exotic profiles.
Notably, there also exists the possibility of spectral line narrowing--the spectral line appears narrower than the telescope should conventionally be capable of resolving.
Aside from experimental error, we are unaware of any other possible cause of this final signature, which could (if observed) constitute strong evidence for quantum nonequilibrium and the deBB theory. If a source of such a nonequilibrium signal were detected and could be reliably measured, a final definitive proof could be arrived at by subjecting the signal to a specifically quantum-mechanical experiment as, for example, is described in Ref.\ \cite{AV04a}.

Our paper is organized as follows. In Sec.~\ref{sec:2} we explain why we have chosen to focus on spectral lines. We summarize what appears to be the essential difference between conventional spectral effects and those caused by quantum nonequilibrium, and why this is particularly relevant for spectral lines. We also provide some background helpful to our analysis. Finally, we present an idealized and parameter-free field-theoretical model of a spectral measurement of the electromagnetic field, which is used to represent a telescope photon detector. In Sec.~\ref{sec:3} we present the pilot-wave description of the model and provide explicit calculations showing the result of introducing quantum nonequilibrium.
In our concluding {Sec.~\ref{sec:4}} we outline the phenomena that we judge the most likely to betray the presence of quantum nonequilibrium in spectral lines, and we discuss possible implications for the indirect search for dark matter with reference to three recent controversial spectral features.

\section{Modeling a telescope x-/$\gamma$-ray photon detector}\label{sec:2}
Relaxation to quantum equilibrium is thought to proceed efficiently for systems with sufficiently complicated quantum states \cite{VW05,TRV12,CS10}. 
For everyday matter therefore, with its long and violent astrophysical history, it is expected that quantum nonequilibrium will have long since decayed away. 
As discussed in Ref.\ \cite{UV15}, however, for more exotic particle species that decouple at very early times there exist windows of opportunity whereby quantum nonequilibrium could have been preserved. 
It is then not inconceivable that dark matter may still exhibit nonequilibrium statistics today, should it take the form of particles that indeed decoupled very early.
The search for dark matter is therefore a natural choice of context for the discussion of relic quantum nonequilibrium. For the reasons detailed below, we focus our discussion on the search for a smoking-gun spectral line by x-/$\gamma$-ray space telescopes.
Such a line could, for example, be produced in the $\gamma$-ray range by the $XX\rightarrow\gamma\gamma$ annihilation of various weakly interacting massive particle (WIMP) dark matter candidates \cite{BS88,Rudaz89,B97,B04,Creview16} or in the x-ray range by the decay of sterile neutrinos \cite{DW94,Abazajian01}.  Recent searches have been carried out in the $\gamma$-ray range by Refs.\ \cite{fermi3.7,fermi5.8,Pull07,Mack08,W12,SF12-1,Albert14,HESS16line} and in the x-ray range by Refs.\ \cite{Bulbul14,Boyarsky14,A15,Hitomi16}.
 Although dark matter does not interact directly with the electromagnetic field and so these processes are typically suppressed, it has long been argued that the detection of such a line may be among the most promising methods available to discover the nature of dark matter \cite{BS88}. 
Primarily this is because WIMP $XX\rightarrow\gamma\gamma$ annihilation would create two photons of energy $E_\gamma=m_\text{WIMP}$ and single-photon sterile neutrino decay would create photons of energy $E_\gamma=m_s/2$. The resulting spectral lines would hence yield the mass of the dark matter particle in addition to its spatial location. The hypothetical lines would furthermore appear with only very minimal (mostly Doppler) broadening ($\sim$0.1\% of $E_\gamma$) helping them to be distinguished from the background \cite{Berg12,Abazajian01}. 

The search for such dark matter lines is also, arguably, a promising context in which to consider possible signatures of relic quantum nonequilibrium.
As discussed in Ref.\ \cite{UV15}, field interactions have the effect of transferring nonequilibrium from one quantum field to another.
This means that, if a nonequilibrium ensemble of dark matter particles annihilates or decays in the manner described, we may reasonably expect some of the nonequilibrium to be transferred to the created photons.
If these photons subsequently travel to a telescope without scattering significantly, they could retain the nonequilibrium until their arrival at the telescope.  
As a result, we may focus on modeling the interaction of the detector with nonequilibrium photons, rather than considering interactions of (as yet unknown) dark matter particles. 
That the putative photons are in the x-/$\gamma$-ray range is relevant as modern x-/$\gamma$-ray telescopes are capable of single-photon detection--an inherently quantum process.
(In particular, $\gamma$-ray telescope calorimeters are designed more like particle physics experiments than traditional telescopes.)
A model of such a detector need only describe the measurement of individual photons, affording a useful simplification.

\subsection{Role of the energy dispersion function}\label{role_of_D}
The hypothetical photon signals are close to monoenergetic, suffering only $\sim$0.1\% broadening from conventional sources \cite{Berg12,Abazajian01}. This is significant for our discussion, as we now explain.
When a photon of energy $E_\gamma$ arrives at a real telescope, the telescope photon detector may record a range of possible energies. 
These possible energy readings are distributed according to the detector's energy dispersion function, commonly denoted $D(E|E_\gamma)$ (see for example Refs. \cite{LAT12,XMM_MOS01}). 
This is approximately Gaussian, centered on the true energy, and with a spread quantified by the detector's energy resolution $\Delta E/E_\gamma$. (See for instance Sec.~7 and Fig.~67 in Ref.\ \cite{LAT12}.) If a telescope receives photons distributed with a true spectrum $\rho_\text{true}(E_\gamma)$, it observes a spectrum
\begin{align}
\rho_\text{obs}(E)=\int D(E|E_\gamma)\rho_\text{true}(E_\gamma)dE_\gamma\label{eq1}
\end{align}
that is convolved by $D(E|E_\gamma)$.
Conventional spectral effects, such as the $\sim$0.1\% Doppler broadening expected in the annihilation/decay lines, alter the energy of the signal photons and hence the true spectrum $\rho_\text{true}(E_\gamma)$.
This is not true of quantum nonequilibrium however.
In deBB theory, the system configuration does not affect the standard Schr\"{o}dinger evolution of the quantum state \cite{Holl93}. Photons hence arrive at the telescope with the same quantum states (and the same associated energy eigenvalues) as they would have in the absence of quantum nonequilibrium. 
Instead, quantum nonequilibrium affects the statistical outcomes of quantum interactions between the photons and the telescope. Thus, quantum nonequilibrium alters the detector's energy dispersion $D(E|E_\gamma)$ and not $\rho_\text{true}(E_\gamma)$. This is the essential difference between conventional spectral effects and those caused by quantum nonequilibrium. Of course, since both $\rho_\text{true}(E_\gamma)$ and $D(E|E_\gamma)$ enter into the integrand of Eq.\ \eqref{eq1}, both kinds of effects contribute to the observed spectrum $\rho_\text{obs}(E)$. The relative size of these contributions may, however, depend strongly on the context.

To illustrate this last point, consider the observation of a spectral line $\rho_\text{true}(E_\gamma)=\rho_\text{line}(E_\gamma)$ according to Eq.\ \eqref{eq1} in two separate regimes. First, consider a high resolution instrument in which $\Delta E/E_\gamma$ [the width of $D(E|E_\gamma)$] is significantly smaller than the width of the signal line $\rho_\text{line}(E_\gamma)$. In this case it is appropriate to make the approximation $D(E|E_\gamma)\approx \delta(E-E_\gamma)$, and so the observed spectrum \eqref{eq1} closely approximates the true signal spectrum, $\rho_\text{obs}(E)\approx \rho_\text{line}(E)$. Thus a high resolution telescope may resolve the profile of the signal line. In this regime, any moderate (order unity) alterations that quantum nonequilibrium makes to $D(E|E_\gamma)$ are subdominant in the observed spectrum. 
Second, consider a low resolution telescope for which the width $\Delta E/E_\gamma$ of the energy dispersion function $D(E|E_\gamma)$ is significantly larger than the width of the signal line.
For this case the appropriate approximation is instead $\rho_\text{line}(E_\gamma)\approx\delta(E_\gamma-E_\text{line})$, and so the observed spectrum \eqref{eq1} closely approximates the instrument's own energy dispersion distribution, $\rho_\text{obs}(E)\approx D(E|E_\text{line})$.
In this second regime conventional broadening is not resolved, whereas moderate changes in $D(E|E_\text{line})$ (perhaps caused by quantum nonequilibrium) would be directly observed.

We may draw the remarkable conclusion that quantum nonequilibrium will be more evident in telescopes of low energy resolution.
With regards to the hypothetical WIMP and sterile neutrino lines with $\sim$0.1\% conventional (Doppler) broadening mentioned above, currently operational telescopes are within the low resolution range. 
For example, the Fermi Large Area Telescope (LAT) has a resolution of $\Delta E/E_\gamma\sim 10\%$ \cite{LAT12}.\footnote{The earlier (1990s) EGRET $\gamma$-ray telescope had a resolution of $\sim$20\% \cite{EGRET93} and the 2015-operational CALET and DAMPE $\gamma$-ray telescopes achieve $\sim$2\% \cite{CALET,DAMPE}. These are also in the low resolution regime for a possible WIMP line with 0.1\% broadening.} 
Indeed, if quantum nonequilibrium produced moderate, order unity changes in the Fermi-LAT energy dispersion distribution $D(E|E_\gamma)$, these changes would appear $\sim$100 times larger than the expected Doppler broadening of the hypothetical annihilation line in the observed spectrum.

\subsection{Idealized model of a photon detector}

To explore the potential consequences of quantum nonequilibrium, we now introduce an idealized, field-theoretical, and parameter-free model of a telescope photon detector.  
We base our model on the standard deBB pilot-wave description of von Neumann measurements \cite{B52b}, which is more commonly applied to discrete spectra. 
As we explain, when applied to a continuous (energy) spectrum, a dispersion distribution $D(E|E_\gamma)$ and a resolution $\Delta E/E_\gamma$ naturally arise.
To model a photon detector, we model a measurement of the electromagnetic field, in which we assume only one photon is present at a time. 
To avoid complications associated with the localizability of photons, we take the electromagnetic field to be quantized within a region that loosely corresponds to the body of the instrument.
For each photon, a single energy is recorded.
Over the course of the experiment, many such readings are taken and the resulting set of values may be compared with model spectra for the purposes of hypothesis testing. 

For the measurement of an observable $\mathcal{A}$ with a discrete and nondegenerate spectrum, a standard deBB measurement proceeds as follows. A system with wave function $\psi(q)$ is coupled to a pointer with wave function $\phi(y)$. A commonly used interaction Hamiltonian is 
\begin{align}\label{basic_interaction}
H_\text{I}=g\mathcal{A}p_y,
\end{align}
where $p_y$ is the conjugate momentum operator of the pointer and $g$ is a coupling constant. The measurement process is taken to begin at $t=0$ and, prior to this, the coupling constant $g$ is taken to be 0.
Thereafter, $g$ is taken to be large enough to ensure that the subsequent evolution is dominated by the interaction Hamiltonian.
With this stipulation, the Schr\"{o}dinger equation takes the simple form
\begin{align}
\partial_t\Psi=-g\mathcal{A}\partial_y\Psi
\end{align}
for the duration of the measurement.
Since the spectrum of $\mathcal{A}$ is assumed discrete and nondegenerate, the system wave function may be decomposed as $\psi(q)=\sum_{n}c_n\psi_n(q)$, where $c_n$ are arbitrary coefficients and $\psi_n(q)$ are eigenstates of $\mathcal{A}$ with the eigenvalues $a_n$. The system evolves as
\begin{align}
\psi(q)\phi(y)\rightarrow \sum_n c_n \psi_n(q)\phi(y-ga_nt).\label{eq:evo}
\end{align}
The outcome probability of the experiment is determined by the effective distribution of the pointer $y$, the marginal Born distribution (hereby called the measured distribution), $\rho_{\text{meas}}(y):=\int |\Psi(q,y)|^2dq$.
Initially the different terms in the sum \eqref{eq:evo} overlap in configuration space, producing interference in the measured distribution. 
Over time each component moves at a speed proportional to its eigenvalue $a_n$. 
If the pointer is prepared in a localized state (perhaps a Gaussian), then after a sufficient time (deemed the duration of the measurement) the measured distribution becomes a sum of nonoverlapping terms,
\begin{align}\label{disjoint}
\rho_{\text{meas}}(y)=\sum_n|c_n|^2|\phi(y-ga_nt)|^2,
\end{align}
and no longer exhibits interference.
Hence, at the end of each measurement, the pointer is found in one of a number of disjoint regions that correspond to the nonoverlapping terms in Eq.\ \eqref{disjoint}.
In the absence of quantum nonequilibrium, there is a $|c_m|^2$ chance of finding the pointer in the $m$th region.
If a single measurement concludes with the pointer in the $m$th region, this implies a measurement outcome of the corresponding eigenvalue, $a_m$.
The discrete spectrum, $|c_n|^2$, may then be reconstructed by repeated measurements over an ensemble.

According to textbook quantum mechanics, quantum state collapse occurs at the end of each measurement in order to ensure that the pointer is found in a single one of the disjoint regions. 
In the deBB account, the system (which always occupies a definite position in configuration space) is simply found in one of the regions, with no need for any nonunitary evolution.
Instead an ``effective collapse'' occurs as, once the components of the wave function \eqref{eq:evo} have properly separated, subsequent evolution of the system configuration is determined solely by the component that contains the configuration. 

This formulation accounts only for the measurement of observables with discrete spectra. One cannot associate disjoint regions in $y$ to eigenvalues on a continuous (energy) scale. Instead, given a particular pointer position, the energy of the incident photon must be estimated. The model photon detector measures the total (normal-ordered) Hamiltonian of the free-space electromagnetic field $\mathopen{:}H_{\text{EM}}\mathclose{:}$, so that in place of Eq.\ \eqref{basic_interaction} the interaction Hamiltonian is
\begin{align}
H_\text{I}=g\mathopen{:}H_{\text{EM}}\mathclose{:} p_y.
\end{align}
For an initial single-photon state $\ket{E_\gamma}$, the state evolution becomes simply
\begin{align}
\ket{E_\gamma}\ket{\phi(y)}\rightarrow \ket{E_\gamma}\ket{\phi(y-gE_\gamma t)}.\label{eq:state_vector}
\end{align}
Although the photon has an exact energy, the quantum uncertainty in the initial position of the pointer produces uncertainty in the pointer position at any later time.
The probability density of finding the pointer at a position $y$ is given by
\begin{align}
\rho_\text{meas}(y,t)=|\phi(y-gE_\gamma t)|^2.\label{eq:meas_point}
\end{align}
If this measured distribution were known, then one could correctly infer the true photon energy $E_\gamma$. 
In a single measurement, however, the pointer is found at a single position that is distributed as $\rho_\text{meas}(y,t)$.
The best estimate of the true energy is that which was most likely to have caused the observed pointer position. 
Taking the initial pointer wave packet $|\phi(y)|^2$ to be Gaussian and centered at $y=0$, this amounts to assigning the energy
\begin{align}
E=y/gt.\label{y_to_E}
\end{align}
The energy dispersion function $D(E|E_\gamma)$ is the distribution of possible energy values recorded by the instrument given the true energy $E_\gamma$. If the pointer packet has initial variance $\sigma_y^2$, the distribution of measured pointer positions \eqref{eq:meas_point} may be translated into the distribution of recorded energies \eqref{y_to_E}, giving
\begin{align}
D(E|E_\gamma) = \frac{1}{\sqrt{2\pi}}\frac{gt}{\sigma_y}e^{-\frac{1}{2}\left(\frac{gt}{\sigma_y}\right)^2(E-E_\gamma)^2}.\label{eq:e_dis_PDF}
\end{align}
Since this is a Gaussian, the energy resolution (the half-width of the fractional 68\% containment window) is simply the fractional standard deviation,
\begin{align}
\frac{\Delta E}{E_\gamma}=\frac{\sigma_y}{gtE_\gamma}.\label{eq:e_dis}
\end{align}
For hypothesis testing, we need to know the spectrum we expect to observe for each potential true spectrum $\rho_\text{true}(E_\gamma)$ . This is given by the convolution \eqref{eq1}, that for our model becomes
\begin{align}
\rho_\text{obs}(E)=\int_0^\infty\rho_\text{true}(E_\gamma)\frac{1}{\sqrt{2\pi}}\frac{gt}{\sigma_y}e^{-\frac{1}{2}\left(\frac{gt}{\sigma_y}\right)^2(E-E_\gamma)^2}\,dE_\gamma,\label{recon_simple}
\end{align}
which is a simple ``Gaussian blur'' (or Weierstrass transform) of the true spectrum.

The duration $t$ of the measurement appears in the denominator of Eq.\ \eqref{eq:e_dis}. Thus, in a sense the precision of the energy reading improves with the run time.
But $t$ appears only as a factor in the quantity $gt$, so a larger coupling constant would also improve the precision.
In Sec.~\ref{sec:3} it is useful to rescale the pointer variable $y$ in terms of its initial standard deviation $\sigma_y$, where it turns out that $\sigma_y$ appears only in the quantity $gt/\sigma_y$.
Thus a narrower pointer packet also improves the precision.
With this in mind, we define the rescaled time variable
\begin{align} 
T= \frac{gtE_\gamma}{\sigma_y}=\left(\frac{\Delta E}{E_\gamma}\right)^{-1},\label{eq:T}
\end{align}
where $\Delta E/E_\gamma=1/T$ is the resolution of the telescope.

The variables $g$, $\sigma_y$, and $t$ are the only free parameters in this model. The definition \eqref{eq:T} allows their replacement with the single easily interpreted quantity $T$.
Thus, for instance, the model may reproduce a roughly EGRET resolution of $20\%$ at $T=5$, a roughly Fermi-LAT resolution of $\sim$10\% at $T=10$, or a roughly CALET/DAMPE resolution of $\sim$2\% at $T=50$.

The true energy $E_\gamma$ is included in the definition \eqref{eq:T} so that the rescaled time $T$ is exactly the reciprocal of the energy resolution \eqref{eq:e_dis}--no matter the true energy of the incident photon. 
Consequently, the effects of quantum nonequilibrium we describe in Sec.~\ref{sec:3} are independent of the energy of the spectral line.

\section{Nonequilibrium spectral lines}\label{sec:3}
As discussed in Sec.~\ref{sec:2}, we consider a low resolution telescope in which the detector energy resolution $\Delta E/E_\gamma$ is significantly larger than the width of the spectral line. 
The signal photons may then be taken to be approximately monoenergetic, $\rho_\text{true}(E_\gamma)=\delta(E_\gamma-E_\text{line})$, and hence the telescope is expected to record photon energies distributed according to the detector energy dispersion function at the line energy, $\rho_\text{obs}(E)=D(E|E_\text{line})$.
In the presence of quantum nonequilibrium, we expect to observe deviations from $D(E|E_\text{line})$.
\subsection{De Broglie-Bohm description of model}
To calculate how these deviations may appear in practice, we now provide a deBB description of the photon detector model introduced in Sec.~\ref{sec:2}. For this we require a coordinate representation of the electromagnetic field. We work in the Coulomb gauge, $\nabla\cdot \mathbf{A}(\mathbf{x},t)=0$, with the field expansion
\begin{align}
\mathbf{A}(\mathbf{x},t)=\sum_{\mathbf{k}s}\left[A_{\mathbf{k}s}(t)\mathbf{u}_{\mathbf{k}s}(\mathbf{x})+A_{\mathbf{k}s}^*(t)\mathbf{u}_{\mathbf{k}s}^*(\mathbf{x})\right],\label{u_expansion}
\end{align}
where the mode functions
\begin{align}
\mathbf{u}_{\mathbf{k}s}(\mathbf{x})=\frac{\bm{\varepsilon}_{\mathbf{k}s}}{\sqrt{2\varepsilon_0 V}}e^{i\mathbf{k.x}}
\end{align}
and their complex conjugates define a basis, and $V$ is a normalization volume that may be thought to correspond loosely to the volume of the instrument.
To avoid duplication of basis elements $\mathbf{u}^*_{\mathbf{k}s}$ with $\mathbf{u}_{\mathbf{-k}s}$, the summation \eqref{u_expansion} should be understood to extend over only half the possible wave vectors $\mathbf{k}$. (See for instance Ref.\ \cite{schiff}.)
This expansion allows the energy of the electromagnetic field to be written as
\begin{align}
U&=\frac12 \int_V\mathrm{d}^3x\left(\varepsilon_0 \mathbf{E}^2+\frac{1}{\mu_0}\mathbf{B}^2\right)\\
&=\sum_{\mathbf{k}s}\frac{1}{2}\left(\dot{A}_{\mathbf{k}s}\dot{A}_{\mathbf{k}s}^*+\omega_\mathbf{k}^2A_{\mathbf{k}s}A_{\mathbf{k}s}^*\right),\label{comp_HOs}
\end{align}
where $\omega_\mathbf{k}=c|\mathbf{k}|$.
Equation \eqref{comp_HOs} corresponds to a decoupled set of complex harmonic oscillators of unit mass. We prefer instead to work with real variables and so decompose $A_{\mathbf{k}s}$ into its real and imaginary parts 
\begin{align}
A_{\mathbf{k}s}=q_{\mathbf{k}s1} +iq_{\mathbf{k}s2}.
\end{align}
The free field Hamiltonian may then be written as
\begin{align}
H_0=\sum_{\mathbf{k}sr}H_{\mathbf{k}sr}\label{ham_decoupled}
\end{align}
with $r=1,2$, where
\begin{align}
H_{\mathbf{k}sr}=\frac{1}{2}\left(p_{\mathbf{k}sr}^2+\omega_{\mathbf{k}}^2q_{\mathbf{k}sr}^2\right)
\end{align}
and where $p_{\mathbf{k}sr}$ is the momentum conjugate to $q_{\mathbf{k}sr}$.

The variables $q_{\mathbf{k}sr}$ and $y$ are the configuration space ``beables.'' Together they specify the configuration of the field and pointer. By rescaling the beable coordinates,
\begin{align}
Q_{\mathbf{k}sr}&=\sqrt{\frac{\omega_{\mathbf{k}}}{\hbar}}q_{\mathbf{k}sr},\quad Y=\frac{y}{\sigma_y},
\end{align}
and using the rescaled time parameter \eqref{eq:T}, the Schr\"{o}dinger equation may be written as\footnote{In Eqs.\ \eqref{schro}--\eqref{gen_guidance_2}, $E_\gamma$ should be understood to be a reference energy that later refers to the energy of the incident photon.}
\begin{align}
\partial_T\Psi +\frac12 \sum_{\mathbf{k}sr}\frac{E_\mathbf{k}}{E_\gamma}\left(-\partial_{Q_{\mathbf{k}sr}}^2 + Q_{\mathbf{k}sr}^2-1\right)\partial_Y\Psi=0.\label{schro}
\end{align}
The following deBB guidance equations\footnote{Note that in deBB, one of two equivalent formalisms may be adopted. We use the first-order formalism in which guidance equations specify the ``velocity'' of the beables in configuration space. Some authors prefer the second-order pseudo-Newtonian formalism in which the acceleration and initial velocity are instead specified. This allows the definition of a so-called quantum potential energy, but as this is not conserved it is not often of much practical use. In the following, references to energy refer only to the standard notion. For further discussion see Refs.\ \cite{Holl93,CV14}.} may then be derived by using a similar method to that used in \cite{UV15} (based on general expressions derived in \cite{SV08}),
\begin{align}
\partial_T Q_{\mathbf{k}sr}=&\frac{E_\mathbf{k}}{E_\gamma}\left(-\frac13 \Psi\partial_{Q_{\mathbf{k}sr}}\partial_Y \Psi^* +\frac16 \partial_Y\Psi\partial_{Q_{\mathbf{k}sr}}\Psi^*\right.\nonumber\\
&\left.+\frac16 \partial_{Q_{\mathbf{k}sr}}\Psi\partial_Y\Psi^* -\frac13 \Psi^*\partial_{Q_{\mathbf{k}sr}}^2\Psi \right)/|\Psi|^2,\label{gen_guidance_1}\\
\partial_TY=&\sum_{\mathbf{k}sr}\frac{E_\mathbf{k}}{E_\gamma}\left(-\frac16 \Psi\partial_{Q_{\mathbf{k}sr}}^2\Psi^*+\frac16 \partial_{Q_{\mathbf{k}sr}}\Psi\partial_{Q_{\mathbf{k}sr}}\Psi^*\right.\nonumber\\
&\left.-\frac16 \Psi^*\partial_{Q_{\mathbf{k}sr}}^2\Psi + \frac12 \left(Q_{\mathbf{k}sr}^2 -1\right)|\Psi|^2\right)/|\Psi|^2.\label{gen_guidance_2}
\end{align}
These are the equations of motion for the $(\{Q_{\mathbf{k}sr}\},Y)$ configuration under a general quantum state. 
Prior to the spectral measurement, one mode of the field contains a nonequilibrium photon of energy $E_\gamma$. The beable associated with this photon-carrying mode is referred to as $Q$.
Henceforth, all summations or products over $\mathbf{k}sr$ should be understood to exclude the mode that contains the photon.
With this in mind, the wave function(al) of the pointer-field system may be written as
\begin{align}
\Psi=&\underbrace{(2\pi)^{-\frac14}\exp\left[-\frac14\left(Y-T\right)^2\right]}_{\phi}\nonumber\\
&\times\underbrace{2^{\frac12}\pi^{-\frac14}Q\exp\left[-\frac12 Q^2\right]}_{\chi_1}
 \prod_{\mathbf{k}sr}\underbrace{\pi^{-\frac14} \exp\left[-\frac12 Q_{\mathbf{k}sr}^2\right]}_{\chi_0}.\label{wvfn_all_modes}
\end{align}
Here, $\phi$ is the pointer packet while $\chi_0$ and $\chi_1$ refer to harmonic oscillator ground and first excited states respectively. For this specific state the guidance Eqs.\ \eqref{gen_guidance_1} and \eqref{gen_guidance_2} become
\begin{align}
\partial_TQ&=\frac16\left(\frac1Q-Q\right)(Y-T),\nonumber\\
\partial_TQ_{\mathbf{k}sr}&=-\frac16 \frac{E_{\mathbf{k}}}{E_\gamma}Q_{\mathbf{k}sr}(Y-T),\nonumber\\
\partial_TY&=\frac{1}{6Q^2}+\frac13 Q^2 +\frac16 +\sum_{\mathbf{k}sr}\frac{E_\mathbf{k}}{E_\gamma}\left(\frac13 Q_{\mathbf{k}sr}^2 -\frac16\right)\label{spec_guidance_full}.
\end{align}

The pointer is coupled to the total energy of the field.
Quantum mechanically the vacuum modes are effectively uncoupled from the pointer, as is evident from the simple Schr\"{o}dinger evolution \eqref{wvfn_all_modes}.
But in the deBB treatment the guidance Eqs.~\eqref{spec_guidance_full} describe a system in which the beables of each vacuum mode, $Q_{\mathbf{k}sr}$, are coupled directly to the pointer, and through their interaction with the pointer they are coupled indirectly to each other.
(For more details on the energy measurement of a vacuum mode see \cite{UV15}.)
This is accordingly a very complex high-dimensional system. Since our purpose is not to provide an accurate description of a real telescope photon detector, but merely to provide illustrative, qualitative examples of possible phenomena, we now truncate the model.

As a first approximation to the full deBB model of Eqs.\ \eqref{spec_guidance_full}, we consider a system in which the pointer beable is decoupled from the vacuum mode beables. In this reduced system, only the beable $Q$ of the excited mode affects the evolution of the pointer, and so the system is effectively two dimensional. 

\subsection{Length scale of nonequilibrium spectral anomalies}
\begin{figure}
\includegraphics{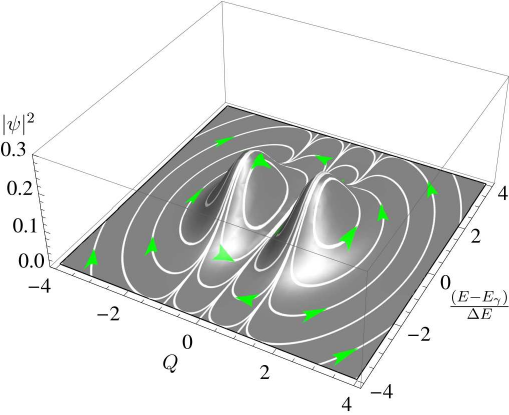}
\caption{\label{fig:trajectories}Periodic orbits in the configuration [Eq.\ \eqref{paths}] produced by the guidance Eqs.\ \eqref{eq:reduced_guidance} contrasted with the quantum equilibrium distribution \eqref{reduced_equilibrium}.}
\end{figure}
Rather than using the variable $Y$ to track the evolution of the pointer, it is convenient to translate this directly into an energy reading. To this end, define the variable $\text{dev}E:=\left(E-E_\gamma\right)/\Delta E$. This is the deviation from a perfect energy reading in units of the detector energy resolution $\Delta E$, and is related to $Y$ by $\text{dev}E=Y-T$. The truncated configuration space may then be spanned by coordinates $(Q,\text{dev}E)$. In terms of these coordinates, the quantum equilibrium distribution
\begin{align}\label{reduced_equilibrium}
|\Psi|^2= \frac{1}{\sqrt{2\pi}}\exp\left(-\frac12 \text{dev}E^2\right)\frac{2}{\sqrt{\pi}}Q^2\exp\left(-Q^2\right),
\end{align}
and the guidance Eqs.\ 
\begin{align}
\partial_TQ=&\frac{1}{6}\text{dev}E\left(\frac1Q -Q\right),\nonumber\\
\partial_T\text{dev}E=&\frac{1}{6Q^2}+\frac{1}{3}Q^2-\frac{5}{6},\label{eq:reduced_guidance}
\end{align}
are both time independent. 
The guidance equations specify stationary orbits around points at $(\pm\sqrt{5\pm\sqrt{17}}/2,0)$, with paths given by
\begin{align}
\text{constant}=\text{dev}E^{2}/2+Q^2-\ln|Q|-\ln|Q^2-1|.\label{paths}
\end{align}
These orbits are contrasted with quantum equilibrium in Fig.~\ref{fig:trajectories}. 
Each of the orbits corresponds to an energy reading that oscillates between an overestimation and an underestimation of the true photon energy $E_\gamma$.  
The extent of these oscillations is stationary with respect to the energy scale $\Delta E$ (the detector energy resolution), rather than any fixed energy scale. 
If considered on a fixed energy scale, the oscillations would appear to shrink as the system evolves and the model, in effect, describes the outcome of an increasingly high resolution telescope [according to Eq.~\eqref{eq:T}].  
As a result, the model reflects the discussion in Sec.~\ref{role_of_D}--spectral anomalies caused by quantum nonequilibrium should be expected to be observed on the scale of the telescope energy resolution, rather than any scale independent of the telescope.

\begin{figure*}
\includegraphics{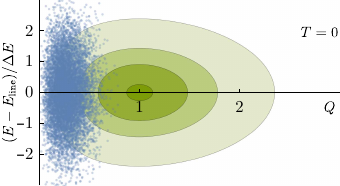}\hspace{3pt}
\includegraphics{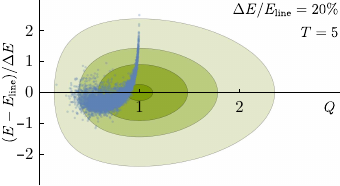}\hspace{3pt}
\includegraphics{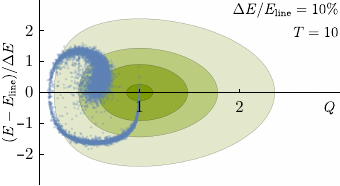}
\\ \vspace*{6pt}
\includegraphics{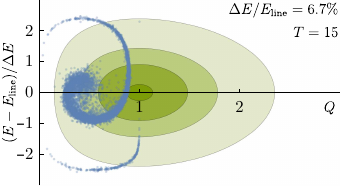}\hspace{3pt}
\includegraphics{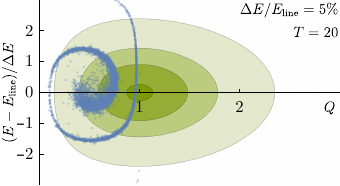}\hspace{3pt}
\includegraphics{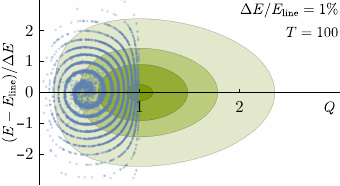}
\caption{\label{w_quarter_relaxation}The evolution of a nonequilibrium ensemble of model spectral measurements (represented by 10,000 points). The individual trajectories that compose the nonequilibrium distribution are displayed in blue and the equilibrium distribution is displayed in green. Only $Q>0$ is shown as the behavior is identical for $Q<0$. Each frame displays the state at particular time $T$ in the evolution. Since $\Delta E/E_\text{line}=T^{-1}$, each frame is also the final system state produced at a particular energy resolution.
The corresponding spectral line profiles (at the corresponding telescope energy resolutions) are shown in Fig.~\ref{w_quarter_marginal}.  
The initial nonequilibrium distribution for the field mode used in this figure is equal to the equilibrium distribution narrowed by a factor of 4 ($w=1/4$). 
In the model, much of the support of such $w<1$ nonequilibrium distributions is confined to the region $|Q|<1$ with the result of producing sharp, narrowed spectral lines (see Fig.\ \ref{w_quarter_marginal}). Frame 6 shows the formation of the fine structure that is a hallmark of quantum relaxation.}
\end{figure*}
\begin{figure*}
\includegraphics{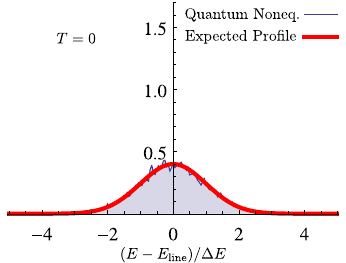}\hspace{-2mm}
\includegraphics{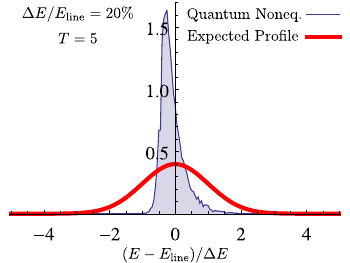}\hspace{-2mm}
\includegraphics{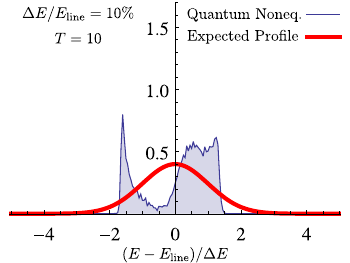}
\\ \vspace*{6pt}
\includegraphics{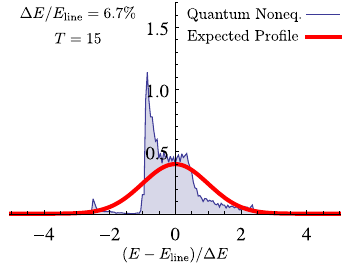}\hspace{-2mm}
\includegraphics{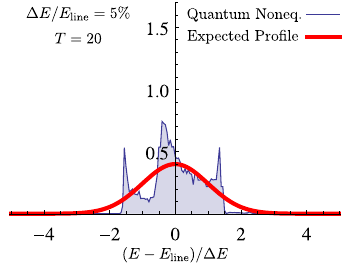}\hspace{-2mm}
\includegraphics{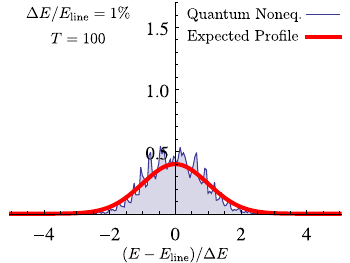}
\caption{\label{w_quarter_marginal}Model spectral line profiles produced by a $w=1/4$ quantum nonequilibrium (corresponding to the frames shown in Fig.~\ref{w_quarter_relaxation}), contrasted with the expected line profile, $D(E|E_\text{line})$. Each frame corresponds to a particular telescope energy resolution $\Delta E/E_\text{line}$. (The third frame, for instance, shows a resolution of $\Delta E/E_\text{line}=10\%$, approximately that of the Fermi-LAT.) The lines are given by the marginal distribution in variable $\text{dev}E:=(E-E_\text{line})/\Delta E$ of the frames in Fig.~\ref{w_quarter_relaxation}. The plots are histograms that have been normalized to represent a probability distribution (plotted on the vertical axis) and hence there is a small amount of statistical fluctuation due to the finite sample size of 10,000. 
Frames 2--4 display distinct signatures of quantum nonequilibrium that could be searched for in experimental data. 
These show features that are too narrow to be conventionally resolved at the corresponding instrument resolution.
Frame 3 also shows a clear double bump that could not be resolved conventionally. These features are commonly produced by the model for $w<1$ nonequilibria.}
\end{figure*}

\begin{figure*}
\includegraphics{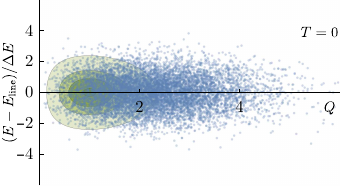}\hspace{3pt}
\includegraphics{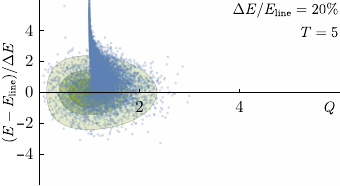}\hspace{3pt}
\includegraphics{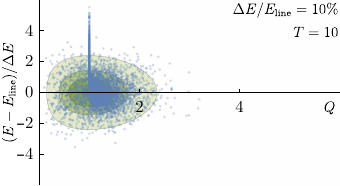}
\\ \vspace*{15pt}
\includegraphics{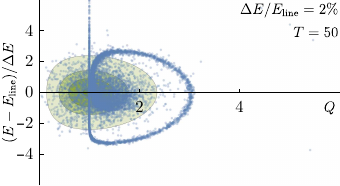}\hspace{3pt}
\includegraphics{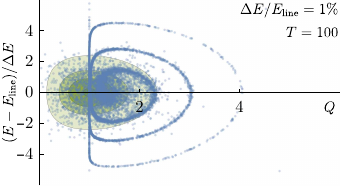}\hspace{3pt}
\includegraphics{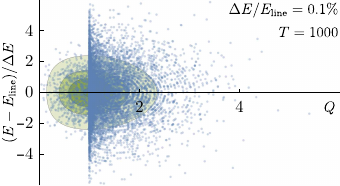}
\caption{\label{fig:w=2_relaxation} The evolution of an ensemble of model spectral measurements (represented by 10,000 points) subject to a $w=2$ quantum nonequilibrium. The individual beable configurations that compose the nonequilibrium ensemble are displayed in blue and the equilibrium distribution is displayed in green. Each frame displays the ensemble at a particular time $T$ in the evolution. Since $\Delta E/E_\text{line}=T^{-1}$, these are also the ensembles found upon the completion of a measurement at the corresponding energy resolution. Only $Q>0$ is shown as the behavior is identical for $Q<0$. In the model, $w>1$ quantum nonequilibrium distributions find more of their bulk confined to the regions $|Q|>1$ (see Fig.~\ref{fig:trajectories}). For lower resolution measurements (frames 2 and 3) this results in an overestimation of the energy and a broadened line. Towards higher resolutions, fine structure develops, leading to a partial quantum relaxation. Spectral line profiles at the corresponding energy resolutions are shown for $w>1$ nonequilibrium in Fig.~\ref{w4_marginal} (we have used $w=4$ for clarity).} 
\end{figure*}

\subsection{Introducing quantum nonequilibrium}
\begin{figure*}
\includegraphics{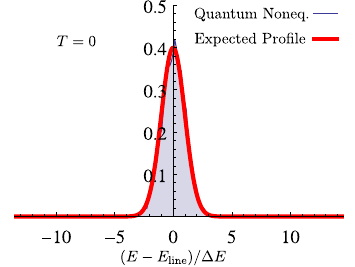}\hspace{-2mm}
\includegraphics{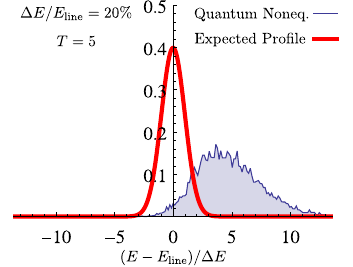}\hspace{-2mm}
\includegraphics{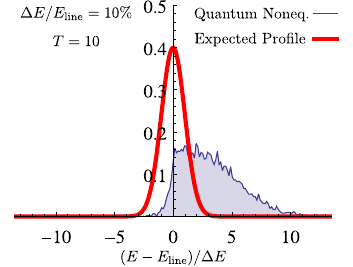}
\\ \vspace*{15pt}
\includegraphics{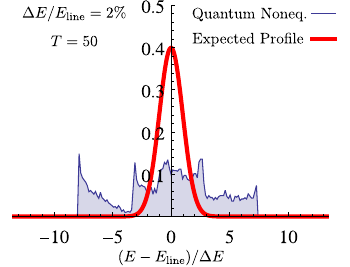}\hspace{-2mm}
\includegraphics{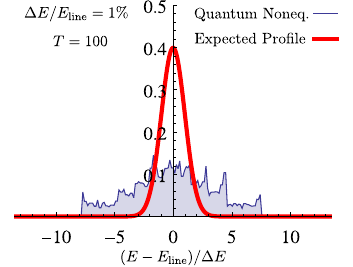}\hspace{-2mm}
\includegraphics{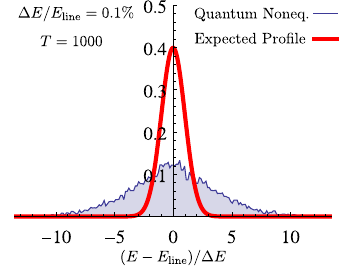}
\caption{\label{w4_marginal}Spectral line profiles produced by a $w=4$ quantum nonequilibrium, contrasted with the expected line profile, $D(E|E_\text{line})$. Each frame corresponds to a model telescope of a particular energy resolution. (The third frame, for instance, shows a line produced by a model telescope with a roughly Fermi-LAT resolution of $\Delta E/E_\text{line}=10\%$.) The horizontal axis denotes the deviation from the true line energy in units of the telescope energy resolution $\Delta E$. (For example, in absolute energy units the expected profile of the spectral line in frame 3 is 100 times the width of that in frame 6.) The plots are histograms that have been normalized to represent a probability distribution (plotted on the vertical axis), and hence there is a small amount of statistical fluctuation due to the finite sample size of 10,000. }
\end{figure*}

As the quantum nonequilibrium evolves (and in effect the system models an increasingly high resolution instrument) we expect to see a dynamic relaxation that contributes transients.
(For an introduction to the mechanics of quantum relaxation, see Refs.\ \cite{AV91a,Uthesis}.)
In order to provide explicit calculations displaying the line profiles produced by quantum nonequilibrium in the photon detector model, we must first specify the initial nonequilibrium photon distributions.
There are, however, as yet no \textit{a priori} indications on the nature of the nonequilibrium distributions that could be present in the photon statistics--the possible or likely shape and extent of the deviations from the Born distribution remain an open question.
With this in mind, in order to provide a simple parameterization of sample nonequilibrium distributions we use a widening parameter $w$ that acts to specify the initial distribution as  
\begin{align}
\rho_0(Q,\text{dev}E)=|\psi_0(Q/w,\text{dev}E)|^2/w.
\end{align}
To investigate the consequences of introducing such photon quantum nonequilibrium into the model detector, ensembles of 10,000 configurations were evolved for each of the widening parameters $w=1/16,1/8,1/4,1/2,2,4,8$ using a standard Runge-Kutta (Cash-Karp) algorithm. 
Representative snapshots of the resulting nonequilibrium distributions and spectral lines are illustrated in \mbox{Figs.~\ref{w_quarter_relaxation}--\ref{w4_marginal}}.

Figure \ref{w_quarter_relaxation} shows snapshots in the evolution of the quantum nonequilibrium for $w=1/4$. (Only the $Q>0$ half of the configuration space is shown as the behavior is identical for $Q<0$.) 
Figure \ref{w_quarter_marginal} contrasts the resulting spectral lines (at the corresponding detector energy resolutions) with the expected line profile, $D(E|E_\text{line})$.

In the model, trajectories are confined to fixed orbits [displayed in Fig.~\eqref{fig:trajectories}] that oscillate between $\text{dev}E>0$ where the photon energy is overestimated and $\text{dev}E<0$ where it is underestimated. Since larger orbits have a larger period, trajectories that begin close to each other (and so initially oscillate in phase) gradually desynchronize. This produces swirling patterns in the configuration space distributions that grow into fine structure, a hallmark of the process of relaxation in classical mechanics and also in deBB quantum theory. 

At relatively low detector energy resolutions ($\Delta E/E_\gamma\gtrsim5\%$) trajectories are still mostly in phase and so mostly grouped together. The corresponding spectral line profiles appear narrower and sharper than could conventionally occur. Since the instrument should not be capable of resolving such a narrow line, lines of this type, if observed, could represent strong evidence for the presence of quantum nonequilibrium. See for instance the second frame of Fig.~\ref{w_quarter_marginal}, which displays such a narrowed line at an energy resolution that approximately matches that of the EGRET instrument. At an energy resolution of $\Delta E/E_\gamma=10\%$, which approximately matches that of the Fermi-LAT, the distribution begins to desynchronize, creating a double line. (See frame 3 of Fig.~\ref{w_quarter_marginal}.) These narrow and double line profiles are common to all initial nonequilibrium distributions that are sufficiently narrow in $Q$. At resolutions beyond about $\Delta E/E_\gamma=5\%$ the model detector produces a line profile that increasingly resembles the standard profile. In principle, observation of the fine structure (that is for instance shown in frame 6 of Fig.~\ref{w_quarter_marginal}) could betray the presence of quantum nonequilibrium, though it is unlikely that this fine structure will be evident in data without a very large sample of readings (signal strength).

Figure \ref{fig:w=2_relaxation} shows snapshots in the evolution when a widened ($w=2$) nonequilibrium photon distribution is introduced. Again, only the $Q>0$ half of the configuration is shown.
Representative spectral line profiles generated by a widened ($w=4$ for clarity) nonequilibrium distribution are shown in Fig.~\ref{w4_marginal}. 
For the $w>1$ cases, more of the bulk of the ensemble density is situated at $|Q|\gtrsim2$. In these regions, the guidance equations \eqref{eq:reduced_guidance} cause early pointer movement in the positive $\text{dev}E$ direction (see Fig.~\ref{fig:trajectories}). Accordingly, low resolution telescope photon detectors produce an overestimated energy. This is shown in frames 2 and 3 of Fig.~\ref{w4_marginal}, although the extent of the overestimations may not be immediately evident. To illustrate, in frame 2 the nonequilibrium line profile is centered on approximately $\text{dev}E=5$ at an approximately EGRET energy resolution of $\Delta E/E_\gamma=20\%$. Consequently, if such a line profile were discovered, it would be likely attributed to an energy of approximately double the correct energy. 

At increasing resolutions (as in the narrowed case) the widened distributions begin to display structure. Double and triple lines appear until at high resolutions the structure becomes fine enough that the line profile approaches an approximately stationary state with respect to the instrument energy resolution. Frame 6 of Fig.~\ref{w4_marginal} displays the profile of the $w=4$ spectral line once it has reached this stage. Note that the amount of line broadening may appear comparable at all resolutions in Fig.~\ref{w4_marginal}, though this is of course only with respect to the energy resolution. For instance, in frame 6 the energy resolution is $0.1\%$, which is equal to the expected conventional broadening so that the effects of quantum nonequilibrium and of conventional broadening would appear approximately the same size. In frame 3, on the other hand, the energy resolution is $10\%$ so that the broadening caused by quantum nonequilibrium is approximately 2 orders of magnitude larger than the conventional broadening.

\section{Discussion and Conclusions}
\label{sec:4}
The de Broglie-Bohm pilot-wave formulation of quantum physics (deBB) allows nonequilibrium violations of the Born rule for quantum probabilities \cite{Uthesis,AV91a,VW05,AV19}. 
A confirmed discovery of this quantum nonequilibrium would serve to distinguish deBB from textbook quantum physics and also from other formulations of quantum theory.
In this paper we have described possible experimental signatures of quantum nonequilibrium in the context of hypothetical dark matter line spectra. 

We have argued in Sec.~\ref{sec:2} that the indirect search for dark matter through spectral lines is an experimental field that is particularly well placed to observe experimental signatures of quantum nonequilibrium (should it exist). 
To support and illustrate our arguments, we have introduced an idealized model of spectral measurement intended to represent an x-/$\gamma$-ray telescope photon detector.
Of course, this model has limited and debatable applicability to actual experiment.
The inner workings of contemporary x-/$\gamma$-ray telescopes are very complicated, typically involving a great many interactions before an estimate of the photon energy may be made.
Our intention, however, is not to provide an accurate description of such telescopes, but only to illustrate spectral features that we argue are to be generally expected in the presence of quantum nonequilibrium. Moreover, since the notion of quantum nonequilibrium is commonly regarded as highly speculative, our intention is to highlight features that are sufficiently exceptional that they might be difficult to account for with conventional physics.

To this end, three key characteristics that may help to betray the presence of quantum nonequilibrium in spectral line signals are as follows.
\paragraph*{\bf{(A) The $\Delta E/E_\gamma$ length scale for line width and structural features.}}
As described in Sec.~\ref{sec:2} and illustrated in Figs.~\ref{w_quarter_marginal} and \ref{w4_marginal}, the effect of quantum nonequilibrium is not to alter the true energy of the signal photons, and hence the true energy spectrum $\rho_\text{true}(E_\gamma)$, but rather to alter the statistics of the interaction of the detector with the photons, and hence the telescope energy dispersion function $D(E|E_\gamma)$. 
The spectral effects of quantum nonequilibrium therefore appear on the length scale of the resolution of the detector ($\Delta E/E_\gamma$) rather than any fixed energy scale as is the case with standard spectral effects.
We regard this as the essential difference between conventional spectral effects and those caused by quantum nonequilibrium.
Telescopes that cannot resolve conventional sources of spectral broadening instead effectively observe their own dispersion function at the line ($\rho_\text{obs}(E)=D(E|E_\text{line})$), and so quantum nonequilibrium produces the largest anomalies in these cases.
(Indeed it is interesting to note that the majority of currently operational x-/$\gamma$-ray telescopes fall into this category when considering the hypothetical 0.1\% Doppler broadened WIMP and sterile neutrino lines that we have used as examples.)
\paragraph*{\bf{(B) Nonequilibrium transients.}}
As illustrated in Sec.~\ref{sec:3}, in the presence of quantum nonequilibrium a dynamical evolution is expected during the measurement process and this contributes deviations from the expected line profile $D(E|E_\text{line})$.
See Figs.~\ref{w_quarter_marginal} and \ref{w4_marginal} for example features.
Taken together, properties A) and B) mean that a number of noteworthy profiles are possible.
Lines may appear narrower than the instrument is canonically capable of resolving (see frames 2 and 3 of Fig.~\ref{w_quarter_marginal}).
Such an occurrence could only otherwise be due to instrumental error or statistical aberration.
It should in practice be possible to exclude these two alternate explanations and so such a signal could represent strong evidence of quantum nonequilibrium.
Lines may appear split into two or three or otherwise display structure (see frames 3--5 of Fig.~\ref{w_quarter_marginal} and frames 3 and 4 of Fig.~\ref{w4_marginal}).
Should it occur, line broadening appears on the length scale of the telescope energy resolution, which is commonly orders of magnitude larger than that expected in the physical energy spectrum (see frames 2 and 3 of Fig.~\ref{w4_marginal}).
For this reason, narrow lines could be mistaken for broad sources, confused with other nearby features or simply lost in the background.

Such deviations will be maximal in telescopes that do not allow for significant relaxation to take place.
In the model this corresponded to lower energy resolution.
It is possible that the many interactions that take place in real telescopes serve to degrade nonequilibrium entirely, making it effectively unobservable. 
However quantum relaxation in many dimensional systems is currently poorly understood and to discount this possibility may be difficult in practice due to the complexity of the model required.
Investigations that help clarify the action of quantum relaxation in the contexts of many body systems and scattering processes would be of great value in this regard.

\paragraph*{\bf{(C) The line shape differs from instrument to instrument.}}
As quantum nonequilibrium affects the dynamical interaction between photon and instrument, the observed line profile is as much a property of the telescope photon detector as it is a property of the nonequilibrium.
As such, two telescopes with large sets of otherwise reliable data may disagree entirely on the profile (or existence) of a line.
While at first glance this may seem to be a blow to the prospect of experimental reproducibility, it could in fact prove quite advantageous.  
For if different telescopes consistently produced differing line profiles, and if these signals were sufficiently well analyzed to ensure no conventional explanations account for the difference, then this could represent substantial evidence for a departure from quantum equilibrium.

In the absence of any nonequilibrium distributions that are more extreme than those we have considered, it seems highly likely that the confirmed discovery of a dark matter spectral line would need to precede any investigation into the cause of an anomalous line shape. 
After all, signal statistics that are significant enough to prove a spectral line has an anomalous shape would surely be significant enough to prove the existence of the line in the first place. 
Nevertheless, the presence of quantum nonequilibrium could have important consequences for the indirect search for dark matter, and this is particularly true for telescopes of lower spectral resolution, where nonequilibrium could obfuscate the detection process, producing ambiguous and anomalous results, making the discovery of a spectral line more elusive. 

In recent years there have been a number of claims of line detections (all controversial), and these may provide context for this last point.
One such controversial line was reported in the galactic dark matter halo at $\sim$133 GeV in Fermi-LAT data in 2012 by Refs.\ \cite{Bring12,W12}. This was originally thought to be a possible WIMP annihilation line \cite{W12}, although its existence is now widely discredited. Indications of instrumental error have been found \cite{Creview16,fermi3.7}, searches with access to more LAT data have reported reduced significance \cite{fermi3.7,fermi5.8}, and a recent search with the ground-based H.E.S.S.\ II Cherenkov telescope ruled out the line with 95\% confidence \cite{HESS16line}.
It is interesting to note, however, that one of the original reasons why this feature was discredited as an actual signal was that it seemed too narrow.
In 2013 the Fermi collaboration disputed the existence of the line, arguing that ``the feature is narrower than the LAT energy resolution at the level of 2 to 3 standard deviations, which somewhat disfavors the interpretation of the 133 GeV feature as a real WIMP signal'' \cite{fermi3.7}.
As we have seen, quantum nonequilibrium could result in features which are narrower than the telescope could canonically be capable of resolving.
Indeed, in our model this is commonly the case for telescope energy resolutions near 10\% (frames 2--4, Fig.~\ref{w_quarter_marginal}).
For such cases, our model also tends to produce multiple lines at these resolutions (frame 3, Fig.~\ref{w_quarter_marginal}) and, remarkably, two prominent early proponents of the feature found it to be marginally better fitted by two lines at $\sim$111 and $\sim$129 GeV (Refs.\ \cite{SF12-1} and \cite{SF12-2}).

A more recent (and yet to be resolved) controversy surrounds a 3.5 keV linelike feature in data from \mbox{x-ray} telescopes. 
The bulk of the original study that reported the line (Ref.\ \cite{Bulbul14}) concerned the stacked spectrum of a large number of nearby galaxies and galaxy clusters\footnote{The advantages of a stacked study are twofold.
The available sample size is enlarged and, as data from many different redshifts are combined in the rest frame of the source, spectral features with an instrumental origin are smeared over while features with a physical origin are superposed; instrumental error is suppressed and physical features are emphasized.} with the CCD instruments aboard the XXM-Newton satellite.
In this stacked study, Ref.\ \cite{Bulbul14} found a faint line (a 1\% bump above the background continuum) with an unresolved profile, consistent with the decay of a 7 keV sterile neutrino. 
Like $\gamma$-ray calorimeters, the CCD instruments employed are capable of single-photon energy measurements, albeit at a relatively low $\sim$100 eV full-width-half-maximum (FWHM) resolution in this range.
This is considerably larger than the $\sim$$0.1\%\times3.5$ keV width of the hypothetical sterile neutrino line, and so the effects of quantum nonequilibrium could also be significant in this context.
While the authors reported the discovery with relatively large ($>3\sigma$) significance, they also noted some apparent anomalies in their data.
Chief among these was the fact that the two separate XXM-Newton CCD instruments used for the stacked study (the MOS and the pn) found best-fit energies at $3.57\pm0.02$ and $3.51\pm0.03$ keV, respectively, a disagreement at a 2.8$\sigma$ level \cite{Bulbul14}.
Also, the best-fit flux from the Perseus Cluster appeared anomalously large in data sets from both XMM-Newton and Chandra instruments.
(This high Perseus flux was confirmed by an independent study using different XMM-Newton data \cite{Boyarsky14}.)
Both of these anomalies were however mitigated if the flux of nearby K and Ar atomic transition lines were allowed to vary from their theoretical values, albeit by at least an order of magnitude.
Since then, a number of related studies have taken place using the CCD instruments aboard the XMM-Newton, Chandra and Suzaku satellites, and the Cadmium-Zinc-Telluride instruments aboard the NuStar satellite.
These paint a confusing picture, with varying levels of detection certainty\footnote{Recently, Ref.\ \cite{NME16} reported an 11$\sigma$ detection in NuStar data from the COSMOS and ECDFS empty sky fields, where a signal from the Milky Way dark matter halo is expected.} and notable nondetections as for instance in Refs.\ \cite{A15,MNE14,JP16,R16}.
For comprehensive reviews of these studies see Refs.\ \cite{I16,White_Paper17}.
Much of the uncertainty stems from two nearby K XVIII lines at 3.47 and 3.51 keV \cite{White_Paper17}.
These lines are difficult to distinguish from the putative sterile neutrino line with $\sim$100 eV resolution instruments.
This matter should have been put to rest with the launch of the Hitomi satellite in February 2016, which carried aboard it the high resolution SXS instrument, capable of a resolution of $\Delta E\simeq 5$ eV FWHM \cite{Hitomi16}, close to that of the theoretical $0.1\%\times 3.5$ keV width of the sterile neutrino line. 
Unfortunately, the Hitomi satellite was lost a month into its mission. Before this, however, it did manage to collect some preliminary data on the Perseus cluster \cite{Hitomi_Perseus,Hitomi16}. Hitomi did not detect the 3.5 keV line at the anomalously high flux level detected by Refs.\ \cite{Bulbul14,Boyarsky14}, ruling it out with $>99\%$ confidence \cite{Hitomi16}.
The Hitomi study did not rule out the lower flux level of the stacked sample in Ref. \cite{Bulbul14}, however, so it is still possible this exists. It also gave no explanation of the reason behind the high flux observations in Refs. \cite{Bulbul14,Boyarsky14}.
The next generation of high spectral resolution telescopes (XRISM, Lynx, and Athena) is due to become operational from the early 2020s onwards.

For our purposes, the discourse that has surrounded the 3.5 keV feature raises a number of relevant points. 
For example, since stacked studies combine the notional line at a range of redshifts, and telescope response is dependent upon photon energy, these studies would likely experience an averaging effect in the observable consequences of quantum nonequilibrium (similar to the smearing of instrumental error). 
Furthermore, since quantum nonequilibrium is an ensemble property, it is dependent upon how the ensemble is defined. 
There is no \emph{a priori} reason, for instance, why two galaxy clusters would produce the same line profile. 
If galactic nonequilibrium distributions did indeed differ, then a stacked study would also average over these, further suppressing nonequilibrium signatures. 
In principle then, quantum nonequilibrium could explain why the single Perseus source produced an anomalous signal that was not observed in the stacked sample. 
Since the observed nonequilibrium signatures are dependent upon the instrument used [property (C) above], it could also explain why the MOS and pn instruments disagreed about the line energy in the stacked spectrum and also why the higher resolution Hitomi instrument did not observe the anomalous Perseus signal.

Indeed, since quantum nonequilibrium is consistent with such a wide range of different anomalies,\footnote{Even the history of observations of the 511 keV electron-positron annihilation line in the galactic bulge (whose existence is not controversial but which has at times been argued to be of dark matter origin \cite{BHSC03,HW04}) can be argued to be consistent with quantum nonequilibrium. Although high resolution ($\Delta E\sim2$ keV FWHM) INTEGRAL-SPI observations now place the line at energy $510.954\pm0.075$ keV with a width of $2.37\pm0.25$ keV FWHM \cite{Churazov04}, early observations of the line were conflicting. Low resolution balloon observations of the galactic center taken in the 1970s placed the line at $476\pm24$ keV \cite{Johnson73} and at $530\pm11$ keV \cite{Haymes75}, and contrary reports of the line flux produced suggestions of time-variation \cite{Leventhal82}--all in principle possible as a result quantum nonequilibrium affecting low resolution spectral measurements.} it may never be as credible as competing explanations with more restricted predictions. 
Even the strongest evidence we have described could be explained in terms of instrument error, and would differ from telescope to telescope. 
This is not an insurmountable barrier to the discovery of particle quantum nonequilibrium, however. 
As we have noted, we would expect the discovery of a line to be confirmed before any significant inquiry into the cause of an anomalous profile.
Indeed, for the reasons we have outlined, it may be the case that a confirmed line discovery will be delayed until a telescope that is capable of resolving the line width is available (in order to minimize the effects of nonequilibrium).
Before such a time, if consistently anomalous signals were detected in lower resolution telescopes, the presence of quantum nonequilibrium could begin to be speculated upon and this might inspire a more direct approach to its detection. 
A definitive proof of the existence of quantum nonequilibrium could be arrived at by subjecting the signal photons to a specifically quantum-mechanical experiment.
One suggestion along these lines is to look for deviations from Malus' law in the polarization probabilities of the signal photons \cite{AV04a}.
There are in fact many possible such experiments that could work. Even a simple double slit experiment, for instance, would show anomalous results--a blurring of the interference pattern--in the presence of quantum nonequilibrium. 
If the incoming photons show anomalies under such tests, then there could be little doubt that the Born rule has been violated.

There are clearly many ways in which our effects could manifest in the search for dark matter, and there are many practical reasons why our effects could turn out to be obscured even if they exist. More work remains to be done on these matters.
Only time will tell whether any of the scenarios we have outlined prove to be informative. 

\acknowledgements
This research was funded in part by the John Templeton Foundation under Grant No. 13171.

\bibliographystyle{apsrev4-1_modified}
\bibliography{../citations}
\end{document}